\RequirePackage{fix-cm}
\documentclass[smallextended]{svjour3}       
\smartqed  
\usepackage{graphicx}
\usepackage{amsmath}
\usepackage{nccmath}
\usepackage{amssymb}
\usepackage{enumerate}
\usepackage{enumitem}
\usepackage{lineno}
\usepackage{color}
\usepackage[center]{caption}
\begin{document}

\title{Improvements on``Secure multi-party quantum summation based on
    quantum Fourier transform''}

\author{Cai Zhang \and Mohsen Razavi \and  Zhewei Sun \and Haozhen Situ
}


\institute{C. Zhang \at
    College of Mathematics and Informatics, South China Agricultural University, Guangzhou, 510642, China \\
    School of Electronic and Electrical Engineering, University of Leeds, Leeds, LS2 9JT, UK \\
    Tel.: +44-7517892138\\
    \email{zhangcai.sysu@gmail.com}           
    \and
    M. Razavi \at
    School of Electronic and Electrical Engineering, University of Leeds, Leeds, LS2 9JT, UK \\
    \and 
    Z.W. Sun \at
    School of Computer Engineering, Shenzhen Polytechnic, Shenzhen, 518055, China \\
    Center for Quantum Computing, Peng Cheng Laboratory, Shenzhen 513055, China \\
    \and
    H.Z. Situ\at
    College of Mathematics and Informatics, South China Agricultural University, Guangzhou, 510642, China 
}

\date{Received: date / Accepted: date}

\maketitle

\begin{abstract}
\linenumbers
Recently, a quantum multi-party summation protocol based on the quantum Fourier transform has been proposed [Quantum Inf Process 17: 129, 2018]. The protocol claims to be secure against both outside and participant attacks. However, a closer look reveals that the player in charge of generating the required multi-partite entangled states can launch two kinds of attacks to learn about other parties' private integer strings without being caught. In this paper, we present these attacks, and propose countermeasures to make the protocol secure again. The improved protocol not only can resist these attacks but also remove the need for the quantum Fourier transform and encoding quantum operations by participants.

\end{abstract}
\section{Introduction}
Since the advent of quantum key distribution (QKD) by Bennett and Brassard \cite{BB84}, a variety of other secure quantum communication protocols has emerged. This includes novel techniques for quantum secure direct communication \cite{xu2019controlled}, quantum secret sharing \cite{chen2018cry}, secure multi-party quantum computation \cite{lo97,crepeau02secure,chau00quantum,ben06secure,smith10multi}, remote state preparation \cite{xu2016novel,chen2017controlled}, which will be part of the growing number of applications that can be run on quantum networks \cite{li2016quantum,chen2019quantum}.
 Of particular interest is the secure multi-party quantum summation \cite{H02,HN03,HKW03,DCWZ07}, as a fundamental primitive for secure multi-party quantum computation \cite{lo97,crepeau02secure,chau00quantum,ben06secure,smith10multi}, whose goals are the protection of participants' input privacy and the guarantee of the correctness of the  summation result. 
Quantum summation has potential applications in quantum anonymous ranking \cite{huang2015quantum}, quantum private comparison \cite{sum2015quantum}, and quantum anonymous voting \cite{wang2016self}. Ji et al. \cite{ji2019quantum} have discussed how to implement these applications by using quantum summation protocols. A considerable number of quantum summation protocols have thus been proposed by using various quantum states, including single photons in two degrees of freedom \cite{zhang2014high}, Bell states \cite{liu2017novel,zhang2019quantum}, Greenberger-Horne-Zeilinger (GHZ) states \cite{chen10efficient}, genuinely maximally entangled six-qubit states \cite{zhang2015three}, and multi-partite high-dimensional entangled states \cite{DCWZ07,ji2019quantum,shi2016secure,shi2017quantum,yang2018secure}.

Lately, Yang et al. \cite{yang2018secure} have proposed a quantum summation protocol, which we refer to by YY2018's protocol, hereafter, based on the quantum Fourier transform. It has been claimed that YY2018's protocol is secure against outside and participant attacks. In this work, we show that, although $P_{1}$, who prepares the $d$-level $n$-component entangled states, cannot collude with other dishonest participants, she can launch attacks to learn about other participants' private integer strings without being detected. We present two kinds of such attacks and propose an improved protocol that renders these attacks invalid. In addition, our new protocol removes the requirement for the quantum Fourier transform and encoding quantum operations to be performed by participants \cite{yang2018secure}.

Note that Shi et al. \cite{shi2016secure} also proposed a quantum summation protocol earlier based on the quantum Fourier transform. However, Shi et al.'s and YY2018's protocols are quite different. Shi et al. protocol adopts the circle-type transmission mode while YY2018's protocol adopts the tree-type transmission mode \cite{yang2018secure}. In addition, Shi et al. employ controlled NOT (CNOT) gates and oracle operators, which are not required in Yang et al.'s work. In Shi et al.'s protocol I, as the authors mention, even though the initiator’s private input ($P_1$'s private input) is unconditionally secure against other dishonest parties, the dishonest parties $P_{k-1}$ and $P_{k+1}$ ($k\ne 1$) can collaborate to steal $P_k$'s private input. To overcome this weakness, they design their improved protocol III in their paper, which is secure against the collusive attacks from any less than $n-1$ dishonest parties. 

The rest of this paper is organized as follows. In Section \ref{sec.rev}, we review YY2018's protocol and give a brief analysis of its security.  In Section \ref{sec.showattack}, we present in detail the two attacks that can be attempted by the first player, $P_1$. 
In Section \ref{sec.improv}, we discuss how to detect $P_{1}$'s attacks and propose an improved protocol, followed by its correctness and security analysis. Conclusions are given in Section \ref{sec.con}.

\section{The YY2018's protocol}\label{sec.rev}
Let us begin with a review of YY2018's protocol \cite{yang2018secure}. In their protocol, $P_{1}$ is assumed not to conspire with other participants and the quantum channels are assumed to be ideal without any noise.

In a group of $n$ players, suppose player $P_{i}$, $i=1,2,\ldots,n$ and $n > 2$,  has a private integer string $K_{i}$ of length $N$ in the following form.
\begin{equation}
\begin{array}{c}
K_{i}=\left(k_{i}^{1}, k_{i}^{2}, \ldots, k_{i}^{N}\right),
\end{array}
\end{equation}
where $k_{i}^{t} \in\{0,1, \ldots, d-1\}$, for $t=1,2, \dots, N$. 
${P}_{1}, {P}_{2}, \ldots, {P}_{n}$ intend to jointly compute the summation of their private integer strings, $K$, given by
\begin{equation}
\begin{array}{lll}
K& = &K_{1} \oplus K_{2} \oplus \ldots \oplus K_{n} \\
 & = &\left(k_{1}^{1} \oplus k_{2}^{1} \oplus \ldots \oplus k_{n}^{1}, k_{1}^{2} \oplus k_{2}^{2} \oplus \ldots \oplus k_{n}^{2}, \ldots, \oplus k_{1}^{N} \oplus k_{2}^{N} \oplus \ldots \oplus k_{n}^{N}\right),
\end{array}
\end{equation}
where $\oplus$ denotes addition modulo $d$.

YY2018's protocol can be described as follows.
\begin{enumerate}[label=(Step {\arabic*})]%
    \item $P_1$ prepares $N$ $d$-level $n$-component entangled states, each of which is  in the state $\frac{1}{\sqrt{d}} \sum_{r=0}^{d-1} | r \rangle_{1}$ $| r \rangle_{2} \ldots | r \rangle_{n}$, and then arranges them into an ordered sequence
    \begin{align}\label{ini}
    \left[\frac{1}{\sqrt{d}} \sum_{r=0}^{d-1} |r\rangle_{1}^{1} | r \rangle_{2}^{1} \ldots | r \rangle_{n}^{1}, \frac{1}{\sqrt{d}} \sum_{r=0}^{d-1} | r \rangle_{1}^{2} | r \rangle_{2}^{2} \ldots | r \rangle_{n}^{2}, \ldots,
    \frac{1}{\sqrt{d}} \sum_{r=0}^{d-1} | r \rangle_{1}^{N} | r \rangle_{2}^{N} \ldots | r \rangle_{n}^{N} \right],
    \end{align}
    where the superscripts $1,2,\ldots,N$ present the order of $d$-level $n$-component entangled states in the sequence. Later, $P_1$ picks up the $i$-th, $i=1,2, \dots, n$, component from each state to construct $n$ sequences labelled as:
    \begin{equation}
    \begin{array}{c}
    S_{1}=\left(p_{1}^{1}, p_{1}^{2}, \ldots, p_{1}^{N}\right), \\
    S_{2}=\left(p_{2}^{1}, p_{2}^{2}, \ldots, p_{2}^{N}\right), \\
    \vdots \\
    S_{n}=\left(p_{n}^{1}, p_{n}^{2}, \ldots, p_{n}^{N}\right),
    \end{array}
    \end{equation}
    where $p_{i}^{t}$, $t=1,2,\ldots,N$, denotes the $i$-th component of the $t$-th entangled state. Next, $P_1$ prepares $n-1$ groups of decoy photons and each photon is randomly chosen from $\{|r\rangle\}_{r=0}^{d-1}$, i.e., the computational basis, or 
    $\{F|r\rangle\}_{r=0}^{d-1}$, i.e., the Fourier basis, where $F$ is the $d$-th order discrete quantum Fourier transform. $P_{1}$ randomly selects one group of the decoy photons and randomly inserts them into 
    $S_j$, $j=2,3,\ldots,n$, to form a new sequence $S^{\prime}_j$. At last, $P_1$ keeps $S_1$ in her hand and sends $S^{\prime}_j$ to $P_j$.
    
    \item \label{step.detection} In this step, $P_{1}$ uses the decoy photons to check if the transmission of $S^{\prime}_j$, $j=2,3,\ldots,n$, is secure with $P_{j}$. If the error rate is higher than a predetermined threshold, they will terminate the protocol; otherwise, they will proceed to the next step.
    
    \item \label{step.enc} $P_j$, $j=2,3,\ldots,n$, discards the decoy photons in $S_{j}^{\prime}$ and obtain $S_j$. $P_j$, $j=1,2,\ldots,n$, encodes her private integer string $K_{j}$ on the components in $S_{j}$ by performing $U_{k^t_{j}}F$, $t=1, 2,\ldots, N$, on component $p_j^{t}$, where $U_{k^t_{j}}=\sum_{u=0}^{d-1} | u \oplus k^t_{j} \rangle\langle u |$ and $\oplus$ represents addition modulo $d$. The new sequence of $S_j$ after encoding is denoted by $ES_{j}$. $P_1$ also encodes her private integer string $K_1$ on the components in $S_1$ by performing $U_{k^{t}_{1}}F$  on component $p^{t}_{1}$. The new sequence of $S_1$ after encoded is denoted by $ES_1$.
    \item \label{step.compute}After all parties have finished encoding their private integer strings, each of them measures all components in their hand in the computational basis $\{|r\rangle\}_{r=0}^{d-1}$ and obtains the corresponding measurement results as follows
    \begin{equation}
    \begin{array}{c}
    M_{1}=\left(m_{1}^{1}, m_{1}^{2}, \ldots, m_{1}^{N}\right), \\
    M_{2}=\left(m_{2}^{1}, m_{2}^{2}, \ldots, m_{2}^{N}\right), \\
    \vdots \\
    M_{n}=\left(m_{n}^{1}, m_{n}^{2}, \ldots, m_{n}^{N}\right),
    \end{array}
    \end{equation}
    where $m_{i}^t$, with $i=1,2,\ldots,n$ and $t=1,2,\ldots,N$, is the measurement result of component $p_{i}^t$  after encoding. $P_{i}$, $i=2,3,\ldots,n$, then announces $M_i$ to $P_{1}$.
     Finally, $P_{1}$ obtains the summation of all parties' private integer strings by computing $M_{1} \oplus M_{2} \oplus \ldots \oplus M_{n}$, which she can then announce.
\end{enumerate}

A quantum summation protocol should be secure against outside as well as participant attacks. In the YY2018's protocol, resilience against outside attacks is achieved by using decoy photons \cite{li2005secure,li2006efficient}, which ensure that the quantum transmission between $P_{1}$ and other parties is secure. 
We find, however, that YY2018's protocol is not necessarily secure against participant attacks. In particular, $P_{1}$ has a chance to carry out effective attacks to achieve other participants' private inputs. Furthermore, we can see that all parties use $U_{k}F$ to encode their private integer strings. This is equivalent to the measurement in the Fourier basis and the addition modulo $d$ of their measurement results and private integer strings. As a result, the quantum operation $U_kF$ can be replaced with the measurement in the Fourier basis and the simple addition modulo $d$ operation. We describe these points in the following sections.

\section{$P_{1}$'s attacks}\label{attack}\label{sec.showattack}
In this section, we show how $P_{1}$ can initiate attacks to obtain other parties' private integer strings without being detected. As we mentioned in Section \ref{sec.rev}, YY2018's protocol in its current form is only able to detect the outside attacks, which makes $P_{1}$'s attacks possible due to her advantage over other parties. $P_{1}$ is able to perform different quantum operations on the states she has prepared, although she cannot conspire with other parties. Here, we describe two kinds of such attacks that enables $P_1$ to steal other parties' private integer strings.
\subsection{Attack 1}
\begin{enumerate}[label=(\arabic*)]
    \item $P_1$ prepares $N$ $d$-level $n$-component entangled states, with each of them being in the state     $\frac{1}{\sqrt{d}} \sum_{r=0}^{d-1} | r \rangle_{1} | r \rangle_{2} \ldots | r \rangle_{n}$,
    and arranges them into an ordered sequence as in equation (\ref{ini}). She then measures each component in the basis $\{|r\rangle\}_{r=0}^{d-1}$. The collapsed states after measurement become
    \begin{equation}
    \left[\left( | r^{1}\right\rangle_{1}, | r^{1}\right\rangle_{2}, \ldots, | r^{1} \rangle_{n} ),\left( | r^{2}\right\rangle_{1}, | r^{2} \rangle_{2}, \ldots, | r^{2} \rangle_{n} ), \ldots,\left( | r^{N}\right\rangle_{1}, | r^{N} \rangle_{2}, \ldots, | r^{N} \rangle_{n} ) ],
    \end{equation}
    where $| r^{t}\rangle_{i}$, with $r^{t} \in \{0,1,\ldots,d-1\}$, $t=1,2,\ldots,N$, and $i = 1,2,\dots,n$, denotes the collapsed state of the $i$-th component in the $t$-th $d$-level $n$-component entangled state after measurement. After that, $P_{1}$ performs $F^{\dagger}$ on each component in the sequence, which results in the following states
    \begin{equation}
    \begin{array}{l}
    \left[( F^{\dagger}| r^{1}\rangle_{1}, F^{\dagger}| r^{1}\right\rangle_{2}, \ldots, F^{\dagger}| r^{1} \rangle_{n} ),\left( F^{\dagger}| r^{2}\right\rangle_{1}, F^{\dagger}| r^{2} \rangle_{2}, \ldots, F^{\dagger}| r^{2} \rangle_{n} ), \\
     \ldots,\left( F^{\dagger}| r^{N}\right\rangle_{1}, F^{\dagger}| r^{N} \rangle_{2}, \ldots, F^{\dagger}| r^{N} \rangle_{n} )],
    \end{array}
    \end{equation}
    where $F$ is the quantum Fourier transform and $F^{\dagger}F=FF^{\dagger}=I$.
    
    Later, $P_{1}$ rearranges the $n$ sequences in the following way:
    \begin{equation}
    \begin{array}{c}
    S_{1} =\left( F^{\dagger}| r^{1}\right\rangle_{1}, F^{\dagger}| r^{2} \rangle_{1}, \ldots, F^{\dagger}| r^{N} \rangle_{1} ), \\ 
    S_{2}=\left( F^{\dagger}| r^{1}\right\rangle_{2}, F^{\dagger}| r^{2} \rangle_{2}, \ldots, F^{\dagger}| r^{N} \rangle_{2} ), \\ 
    \vdots \\
    S_{n} =\left( F^{\dagger}| r^{1}\right\rangle_{n}, F^{\dagger}| r^{2} \rangle_{n}, \ldots, F^{\dagger}| r^{N} \rangle_{n} ). 
    \end{array}
    \end{equation}
    $P_{1}$ then sends $S_j$, $j=2,3,\ldots,n$, to $P_{j}$ using the decoy-photon technique. 
    \item After confirming that $S_j$, $j=2,3,\ldots,n$, has been securely received  by $P_j$, $P_j$ performs $U_{k_{j}^{t}}F$
    on component $F^{\dagger}| r^{t} \rangle_{j}$, $t=1,2,\ldots,N$, as in \ref{step.enc} of the protocol. The corresponding encoded state
     becomes
     \begin{equation}
     (U_{k_{j}^{t}}F)F^{\dagger}| r^{t} \rangle_{j}=U_{k_{j}^{t}}FF^{\dagger}| r^{t} \rangle_{j}=U_{k_{j}^{t}}| r^{t} \rangle_{j}=| k_{j}^{t} \oplus r^{t} \rangle_{j}.
     \end{equation}
     Next, $P_j$ measures all components in the basis $\{|r\rangle\}_{r=0}^{d-1}$, and her measurement results will be
     \begin{equation}
     M_{j}=\left(m_{j}^{1}, m_{j}^{2}, \ldots, m_{j}^{N}\right),
     \end{equation}
     where $m_{j}^{t}=r^{t} \oplus k_{j}^{t}$. $P_j$ then announces $M_j$ to $P_{1}$ in \ref{step.compute} of the protocol. Clearly, $P_{1}$ can easily extract $k_{j}^{t}$ from $m_{j}^{t}$ because she knows $r^{t}$ exactly. $P_{1}$ can therefore obtain other parties' private integer strings without being caught.
\end{enumerate}

\subsection{Attack 2}
In Attack 2, we consider, for simplicity, how $P_{1}$ can steal one integer of one party's (say $P_j$'s, $j \ne 1$) private integer string.
\begin{enumerate}[label=(\arabic*)]
\item $P_1$ prepares one $d$-level $n$-component entangled state, in the state 
$\frac{1}{\sqrt{d}} \sum_{r=0}^{d-1}$ $ | r \rangle_{1} | r \rangle_{2} \ldots | r \rangle_{n}$, and then performs the quantum Fourier transform on each of the last $n-1$ components. The resulting state is given by
\begin{equation}
\frac{1}{\sqrt{d}} \sum_{r=0}^{d-1}| r \rangle_{1} F| r \rangle_{2} \ldots F| r \rangle_{n}.
\end{equation}
She then sends the first component of the above state to $P_{j}$ using the decoy photon technique and holds the remaining $n-1$ components in her hand. 
\item After $P_{j}$ performs $U_{k_j}F$ (assume that one integer of her private integer string is $k_j$) on the received components, the state turns into
\begin{equation}
\begin{aligned}
 &\frac{1}{\sqrt{d}} \sum_{r=0}^{d-1}U_{k_j}F| r \rangle_{1} F| r \rangle_{2} \ldots F| r \rangle_{n} \\
&=d^{-\frac{n-1}{2}} \sum_{l_{1} \oplus l_{2} \oplus\ldots  \oplus l_{n}=0} | l_{1} \oplus k_{j} \rangle \otimes | l_{2}  \rangle \otimes \ldots \otimes | l_{n} \rangle,
\end{aligned}
\end{equation}
where $\oplus$ denotes addition modulo $d$.

 Once $P_j$ measures her components and $P_{1}$ measures the remaining ($n-1$) ones in the basis $\{|r\rangle\}_{r=0}^{d-1}$, the measurement result will be
 \begin{equation}
 | l_{1} \oplus k_{j} \rangle \otimes | l_{2}  \rangle \otimes \ldots \otimes | l_{n} \rangle.
 \end{equation}
Next, $P_j$ announces $| l_{1} \oplus k_{j} \rangle$ to $P_{1}$. In the end, $P_{1}$ can achieve $k_{j}$ by computing 
\begin{equation}
\begin{array}{ll}
&l_{1} \oplus k_{j} \oplus   l_{2} \oplus  \ldots \oplus   l_{n} \\
=&k_{j} \oplus l_{1} \oplus l_{2} \oplus  \ldots \oplus   l_{n} = k_{j}.
\end{array}
\end{equation}
\end{enumerate}
Because the original $n$ sequences are constructed by $P_1$, she could also adopt a similar method for obtaining other parties' private integer strings without being caught. 

Note that, in Attack 2, stealing one integer of each party's private integer string costs one $d$-level $n$-component entangled state. In order to take all other parties' private strings, $N(n-1)$ such entangled states are required, which is more expensive compared with Attack 1 where only $N$ such entangled states are needed. Similarly, $P_1$ can employ other entangled states, such as $d$-level Bell states and $d$-level GHZ states, to get other parties' private integer strings.

\section{The improved protocol}\label{sec.improv}
In the previous section, we showed how $P_{1}$ could succeed in obtaining other parties' private integer strings without being caught. This is mainly because of the lack of a detection mechanism, in YY2018's protocol, to check if the $N$ states shared among $n$ parties are genuinely $d$-level $n$-component entangled states in the form of $\frac{1}{\sqrt{d}} \sum_{r=0}^{d-1} | r \rangle_{1} | r \rangle_{2} \ldots | r \rangle_{n}$. One solution would then be to add a detection step to the protocol before other parties encode their private integer strings on the components sent by $P_{1}$.

Before presenting our method for detecting a dishonest $P_1$, let us explore some of the properties of the main states used in YY2018's protocol. Let $r \in \{0,1,\ldots,d-1\}$ and $d \ge 2$. The set $\{|r\rangle\}_{r=0}^{d-1}$ forms a $d$-level computational basis. The $d$-th order quantum Fourier transform is defined as
\begin{equation}
F | r \rangle=\frac{1}{\sqrt{d}} \sum_{l=0}^{d-1} \zeta^{l r} | l \rangle,
\end{equation}
where $\zeta=e^{2 \pi i / d}$ and the set $\{F|r\rangle\}_{r=0}^{d-1}$ is a $d$-level Fourier basis. 
Expanding each $|r\rangle$, $r=0,1,\ldots,d-1$, in the Fourier basis gives
\begin{equation}
 | r \rangle=\frac{1}{\sqrt{d}} \sum_{l=0}^{d-1} \zeta^{-l r} F| l \rangle.
\end{equation}
The $d$-level $n$-component entangled state, $| \omega \rangle_{12 \ldots n}=\frac{1}{\sqrt{d}} \sum_{r=0}^{d-1} | r \rangle_{1} | r \rangle_{2} \ldots | r \rangle_{n}$, can then be rewritten as
\begin{align}\label{check}
&| \omega \rangle_{12 \ldots n} \nonumber \\ 
&=\frac{1}{\sqrt{d}} \sum_{r=0}^{d-1} | r \rangle_{1} | r \rangle_{2} \ldots | r \rangle_{n}  \nonumber \\ 
&=\frac{1}{\sqrt{d}} \sum_{r=0}^{d-1} (\frac{1}{\sqrt{d}}\sum_{l_{1}=0}^{d-1}\zeta^{-l_{1}r} F|l_{1}\rangle_{1})
(\frac{1}{\sqrt{d}}\sum_{l_{2}=0}^{d-1}\zeta^{-l_{2}r}F|l_{2}\rangle_{2})\ldots (\frac{1}{\sqrt{d}}\sum_{l_{n}=0}^{d-1}\zeta^{-l_{n}r}F|l_{n}\rangle_{n}) \nonumber \\
&=(\frac{1}{\sqrt{d}})^{n+1} \sum_{r=0}^{d-1} \sum_{l_{1},l_{2},\ldots,l_{n}} \zeta^{-r(l_{1}+l_{2}+ \ldots +l_{n})} F|l_{1}\rangle_{1}F|l_{2}\rangle_{2}\ldots F|l_{n}\rangle_{n}) \nonumber \\ 
&=(\frac{1}{\sqrt{d}})^{n-1} \sum_{l_{1} \oplus l_{2} \oplus \ldots \oplus l_{n} = 0} F|l_{1}\rangle_{1}F|l_{2}\rangle_{2}\ldots F|l_{n}\rangle_{n}).
\end{align}
The above state has an interesting property. If each component of $| \omega \rangle_{12 \ldots n}$ is measured in the $d$-level computational basis, then all the measurement results would be the same as, i.e., all subsystems will be in state $|r\rangle$, for a single $r \in \{0,1,\ldots, d-1\}$. If each component of $| \omega \rangle_{12 \ldots n} $ is measured in the $d$-level Fourier basis, however, then the outcome would be in the form $F|l_{1}\rangle_{1}F|l_{2}\rangle_{2}\ldots F|l_{n}\rangle_{n}$, where $l_{1} \oplus l_{2} \oplus \ldots \oplus l_{n} = 0$. We can, therefore, employ this property of $| \omega \rangle_{12 \ldots n}$ to check if it is shared properly among $n$ parties.

We are now ready to explain our improved protocol, whose assumptions are the same as that of YY2018's protocol.  The improved protocol is described in the following.
\begin{enumerate}[label=(S \arabic*)]
    \item $P_{1}$ prepares $N+q$ $d$-level $n$-component entangled sates, with $q$ being a security parameter, each of which is in the state $| \omega \rangle_{12 \ldots n}$. She then chooses the $i$-th, $i=1,2,\ldots,n$, component from each state to construct the following $n$ sequences:
    \begin{equation}
    \begin{array}{c}
    S_{1}=\left(p_{1}^{1}, p_{1}^{2}, \ldots, p_{1}^{N+q}\right), \\
    S_{2}=\left(p_{2}^{1}, p_{2}^{2}, \ldots, p_{2}^{N+q}\right), \\
    \vdots \\
    S_{n}=\left(p_{n}^{1}, p_{n}^{2}, \ldots, p_{n}^{N+q}\right),
    \end{array}
    \end{equation}
    where $p_{i}^{t}$, $i=1,2,\ldots,n$ and $t=1,2,\ldots,N+q$, represents the $i$-th component of the $t$-th entangled state $| \omega \rangle_{12 \ldots n}$. Afterward, $P_{1}$ randomly inserts decoy photons into $S_{j}$, $j=2,3,\ldots,n$, to form a new sequence, denoted by $S^{\prime}_{j}$. Finally, $P_1$ sends $S^{\prime}_{j}$ to $P_j$ and keeps $S_1$ in her hand.
    
    \item 
    Similarly to the \ref{step.detection} of YY2018's protocol, once $P_{j}$ ($j=2,3,\ldots,n$) confirms that she has received $S^{\prime}_{j}$, 
    $P_j$ and $P_1$ utilize the decoy photons to check if the transmission of $S^{\prime}_{j}$ is secure. 
    If the error rate is greater than a predetermined threshold, they will terminate the protocol; 
    otherwise, they will proceed to the next step. 
    
    \item \label{enu.improved.checkgeuine}
    In this step, $P_{j}$, $j=2,3,\ldots,n$, will check if entangled states $| \omega \rangle_{12 \ldots n}$ are genuinely shared among $n$ parties.
    By removing the decoy photons from $S^{\prime}_{j}$, 
    $P_j$ obtains $S_{j}$. 
    Next, $P_{2},P_{3},\ldots,P_{n}$ agree to randomly choose $q$ entangled states $|\omega \rangle_{12 \ldots n}^{\prime}$ from $S_{1},S_{2},\ldots,S_{n}$. 
    For each of the chosen states, they randomly ask $P_{1}$ to measure her component in the computational basis or in the Fourier basis and to announce her measurement result. 
    Other parties also measure their corresponding components in the same basis as $P_{1}$ did. 
    The measurement results, according to equation (\ref{check}), are correlated if all components come from the same entangled state $| \omega \rangle_{12 \ldots n}$ and are measured in the same basis. They compute the error rate relying on these measurement results.  If the error rate is higher than a predetermined threshold, they will terminate the protocol; otherwise, they will proceed to the next step. 
    
    \item By removing $q$ entangled states used for detection, all parties share, with a high probability, $N$ genuine $d$-level $n$-component entangled states $| \omega \rangle_{12 \ldots n}$. All parties measure all the components in their hand in the Fourier basis and get certain measurement results. By mapping $F|r\rangle$, as the output measurement state, to the integer $r$, for $r=0,1,\dots,d-1$, $P_{j}$ ($j=1,2,\ldots,n$) will obtain an integer string $L_{j}=(l_{j}^1,l_{j}^2,\ldots,l_{j}^N)$, with $l_{j}^t \in \{0,1,\ldots,d-1\}$ and $t=1,2,\ldots,N$. After that, $P_{j}$, $j=2,3,\ldots,n$, computes $M_j=K_j\oplus L_j=(k_j^1\oplus l_{j}^{1}, k_j^2\oplus l_{j}^{2} ,\ldots, k_j^N \oplus l_{j}^N )$, and sends it to $P_{1}$. $P_{1}$ also computes $M_1=K_1\oplus L_1$.
    Because $l_{1}^{t}\oplus l_{2}^{t}\oplus \ldots \oplus l_{n}^{t} = 0$, in accordance with equation (\ref{check}), 
    $P_{1}$ can eventually achieve the summation of parties' private integer strings by the following computation
    \begin{equation}\label{final}
    \begin{aligned} &M_{1} \oplus M_{2} \oplus \ldots \oplus M_{n}\\ &=\left(
    \bigoplus\sum_{j=1}^{n}k_j^1\oplus l_{j}^{1}, 
    \bigoplus\sum_{j=1}^{n}k_j^2\oplus l_{j}^{2}, 
    \ldots,
    \bigoplus\sum_{j=1}^{n}k_j^N\oplus l_{j}^{N}\right) \\ 
      &=\left(
      \bigoplus\sum_{j=1}^{n}k_j^1\oplus \bigoplus\sum_{j=1}^{n} l_{j}^{1},
      \bigoplus\sum_{j=1}^{n}k_j^2\oplus \bigoplus\sum_{j=1}^{n} l_{j}^{2},
      \ldots,
      \bigoplus\sum_{j=1}^{n}k_j^N\oplus \bigoplus\sum_{j=1}^{n} l_{j}^{N}
      \right) \\
     &=\left(\bigoplus\sum_{j=1}^{n}k_j^1, \bigoplus\sum_{j=1}^{n}k_j^2, \ldots,\bigoplus\sum_{j=1}^{n}k_j^N\right) \\ &=K_{1} \oplus K_{2} \oplus \ldots \oplus K_{n}=K, \end{aligned}
    \end{equation}
    where $\bigoplus\sum_{i}^{n}x_i=x_1\oplus x_2 \oplus  \ldots \oplus x_n$. $P_1$ then announces the result.
\end{enumerate}

\subsection{Analysis}
Let us move on to the analyses of our improved protocol's correctness and security.

We first analyze the correctness. Note that we also utilize $d$-level $n$-component entangled states $| \omega \rangle_{12 \ldots n}$ to design the improved protocol and that the state $F|r\rangle$ corresponds to the integer $r$, for $r=0,1,\dots,d-1$.
For each entangled state used for the summation of all parties' private integer strings, 
$P_{j}$, $j=1,2,\dots,n$, measures her component in the Fourier basis; if $P_{j}$ obtains the measurement result $F|l_j^{t}\rangle$, $t=1,2,\ldots,N$, she will get the corresponding integer $l_j^{t}$ that has the relation $l_{1}^{t}\oplus l_{2}^{t}\oplus \ldots \oplus l_{n}^{t} = 0$ according to equation (\ref{check}). 
Thus, after receiving $M_j=K_j\oplus L_j=(k_j^1\oplus l_{j}^{1},  k_j^2\oplus l_{j}^{2} ,\ldots, k_j^N \oplus l_{j}^N )$, $j=2,3,\ldots,n$, sent by $P_{j}$, 
$P_1$ can eventually obtain the correct result, as shown in equation (\ref{final}).

For the security, because we also employ decoy photon technique to detect outside attacks, the improved protocol is still secure against all kinds of outside attacks. It also remains secure against attacks from $n-2$ dishonest parties (excluding $P_1$ because she cannot conspire with any dishonest party). Suppose they want to steal $P_{1}$'s and $P_i$'s private integer strings, they have to learn about the exact values of $l_1^t$ and $l_i^{t}$, $t=1,2, \ldots, N$, that are used to encrypt 
$P_{1}$'s and $P_{i}$'s private integer strings, respectively. This, however, is impossible, because $l_1^t$ and $l_i^{t}$ are only known to $P_{1}$ and $P_{i}$, respectively. Consequently, the  ($n-2$)-party collusive attack would not be effective in the improved protocol.

We focus now on $P_{1}$'s attacks. As we can see from equation (\ref{check}), if the genuine entangled state $| \omega \rangle_{12 \ldots n}$ is shared in the same way as proposed in the improved protocol, $P_{1}$ is unable to attain other parties' private integer strings since she does not know the exact values of $l_i^{t}$, $i=2,3,\ldots,n$ and $t=1,2,\ldots,N$. She will try to construct special quantum states to steal other parties' secret, as presented in Section \ref{attack}. However, if $P_{1}$ launches such attacks, she will be caught in Step \ref{enu.improved.checkgeuine} of the improved protocol, as the correlations will be destroyed, according to equation (\ref{check}).

Suppose that $P_{1}$ makes the similar Attack $1$ to obtain some parties' private integer strings, she may prepare fake states as
\begin{equation}
\begin{array}{l}
( F| r\rangle_{1}, F| r\rangle_{2}, \ldots, F| r \rangle_{n} ),
\end{array}
\end{equation}
where $r \in \{0,1,\ldots,d-1\}$. In order to pass the detection with high probability in Step \ref{enu.improved.checkgeuine} of the improved protocol, $P_{1}$ has to prepare the fake states as few as possible.
 
  Let us consider the case where $P_{1}$ only wishes to attain one integer of each party's integer string. She can then prepare only one fake quantum state $( F| r\rangle_{1}, F| r\rangle_{2}, \ldots, F| r \rangle_{n} )$ and $N+q-1$ genuine entangled states $| \omega \rangle_{12 \ldots n}$. In this case, $P_{1}$ has the maximum possibility of passing the detection.
  
  The probability of $P_1$ not being detected by our proposed mechanism can be calculated as follows. Note that the fake quantum state should not be chosen for detection so that  $P_{1}$ can succeed in obtaining one integer of each party's integer string. Since $P_{2},P_{3},\ldots,P_{n}$ randomly choose $q$ entangled states for detection, the probability that $P_{1}$ passes the detection and gets the private integer is
\begin{equation}
    {{N+q-1}\choose{q}} / {{N+q}\choose{q}} = \frac{N}{N+q},
\end{equation}
which will approach to $0$ when $q$ is sufficiently large.

If the fake quantum state is chosen for detection, $P_{1}$ could still pass the detection with probability $(\frac{1}{2}+\frac{1}{2d})$. However, she cannot obtain any information about other parties' private integer strings, as she shares $| \omega \rangle_{12 \ldots n}$ among all the parties and the measurement results of other parties are unknown to her. The analysis of the attacks based on fake entangled states (Attack $2$) is similar. These two kinds of attacks thus become ineffective in the improved protocol.

In fact, once the genuine entangled state $|\omega\rangle_{12\ldots n}$ is shared among $n$ parties, it would be impossible for any group of less than $n-1$ parties (including $P_{1}$) to obtain the remaining two parties' private inputs as their private keys cannot be exactly obtained. Note that in the improved protocol, there is no need for parties to encode their private integer strings. That has been replaced with the measurement in the Fourier basis and classical addition modulo $d$.

We would like to point out that two months after submission of this manuscript, a similar work, by Gu and Hwang, appeared on arXiv \cite{gu2019improvement}. Let us compare Gu and Hwang's work with our work in terms of the attacks, the detection methods, and the improved protocols. In terms of attacks, Gu and Hwang's only attack is almost the same as our Attack 1. The only difference is in Step 1. If $P_1$ in our protocol can also initially prepare $n$ single $d$-level states, with each of which from the basis $\{ F|0\rangle, F|1\rangle,\ldots, F|d-1\rangle\}$, instead of an $n$-component entangled state, then these two attacks are the same. In our case, we also present another attack, Attack 2, based on entangled states, which is not discussed in Gu and Hwang's work. To solve the security problem, Gu and Hwang's work and our work both use the same method to check if the initial state $|\omega\rangle_{1,2,\ldots,n}$ is correctly shared among participants. For the improved protocols, Gu and Hwang's and Yang et al.'s are the same except for the step added to detect $P_{1}$'s dishonesty. In our work, we changed YY2018's protocol a bit more. In our improved protocol, in addition to the new step for $P_{1}$'s dishonesty detection, we also remove the need for the quantum Fourier transform and encoding quantum operations by participants. Instead, noting that the operation $U_k F$ to encode private integer strings is equivalent to the measurements in the Fourier basis followed by addition modulo $d$ of the measurement results, we offer a simpler technique to implement the protocol.

\section{Conclusions}\label{sec.con}
We showed that the protocol in \cite{yang2018secure} could not prevent $P_{1}$, who prepares $d$-level $n$-component entangled states, from obtaining other parties' private integer strings. We analyzed in detail $P_{1}$'s two kinds of attacks and proposed an improved protocol that rectified such security threats. In addition, our improved protocol did not require quantum operations to encode parties' private integer strings. 
\begin{acknowledgements}
This work is supported by the National Natural Science Foundation of China (Grant Nos. 11647140, 61602316, 61502179, 61472452, 61202398), the Natural Science Foundation of Guangdong Province of China (Grant No. 2018A030310147, 2014A030310265), and the Science and Technology Innovation Projects of Shenzhen (No. JCYJ20170818140234295). Mohsen Razavi acknowledges the support of UK EPSRC Grant EP/M013472/1. Cai Zhang is sponsored by the State Scholarship Fund of the China Scholarship Council. All data generated in this paper can be reproduced by the provided methodology.
\end{acknowledgements}


\begin{thebibliography}{}
    %
    %
    \bibitem{BB84}
    Bennett, C.H., Brassard, G.: Quantum cryptography: public-key distribution and coin tossing. In:
    Proceedings of the IEEE International Conference on Computers, Systems and Signal Processing, pp.
    175–179. IEEE Press, Bangalore (1984)
    \bibitem{xu2019controlled}
    Xu, G., Xiao, K., Li, Z.P., et al.: Controlled secure direct communication protocol via the three-qubit partially entangled set of states. CMC-Computers Materials and Continua, 2019, 58 (3): 809-827 
    \bibitem{chen2018cry}
    Chen, X.B., Tang, X., Xu, G., et al.: Cryptanalysis of secret sharing with a single d -level quantum system. Quantum Inf Process 17, 225 (2018)
   \bibitem{lo97}
   Lo, H.K.: Insecurity of quantum secure computations. Phys. Rev. A 56, 1154 (1997)
   \bibitem{crepeau02secure}
    Cr{\'e}peau, C., Gottesman, D., Smith, A.: Secure multi-party quantum computation. In: Proceedings of the thirty-fourth annual ACM symposium on Theory of Computing, pp. 643-652 (2002)
   \bibitem{chau00quantum}
   Chau, H. F.: Quantum-classical complexity-security tradeoff in secure multiparty computations. Phys. Rev. A 61(3), 032308 (2000)
    \bibitem{ben06secure}
    Ben-Or, M., Cr{\'e}peau, C., Gottesman, D., Hassidim, A. Smith, A.: Secure multiparty quantum computation with (only) a strict honest majority. In: Foundations of Computer Science, 2006. FOCS'06. 47th Annual IEEE Symposium on, pp. 249-260 (2006)
   \bibitem{smith10multi}
   Smith, A.: Multi-party Quantum Computation. arXiv:quant-ph/0111030 (2010)
   \bibitem{xu2016novel}
    Xu, G., Chen, X.B., Dou, Z., et al.: Novel criteria for deterministic remote state preparation via the entangled six-qubit state. Entropy 18, 267  (2016) 
    \bibitem{chen2017controlled}
    Chen, X.B., Sun, Y.R., Xu, G., et al.: Controlled bidirectional remote preparation of three-qubit state. Quantum Inf Process 16, 244 (2017)  
    \bibitem{li2016quantum}
    Li, J., Chen, X.B., Sun, X.M., et al.: Quantum network coding for multi-unicast problem based on 2d and 3d cluster states. Sci. China-Inf. Sci. 59, 042301 (2016)                 \bibitem{chen2019quantum}
    Chen, X.B., Wang, Y.L., Xu, G., et al.: Quantum network communication with a novel discrete-time quantum walk. IEEE Access, 7, 13634-13642 (2019)
      \bibitem{H02}
    Heinrich, S.: Quantum summation with an application to integration. Journal of Complexity, 18(1), 1-50 (2002)
    \bibitem{HN03}
    Heinrich, S., Novak, E.: On a problem in quantum summation. Journal of Complexity, 19(1), 1-18 (2003)
    \bibitem{HKW03}
    Heinrich, S., Kwas, M., Wo{\'z}niakowski, H.: Quantum Boolean summation with repetitions in the worst-average setting. In: Monte Carlo and Quasi-Monte Carlo Methods, pp. 243-258. Springer, Heidelberg. (2004)
    \bibitem{DCWZ07}
    Du, J.Z., Chen, X.B., Wen, Q.Y., et al.: Secure multiparty quantum summation. Acta Physica Sinica 56(11), 6214 (2007)
  
    \bibitem{huang2015quantum}
    Huang, W., Wen, Q.Y., Liu, B., et al.: Quantum anonymous ranking. Phys. Rev.A 89(3), 032325 (2014)
    \bibitem{sum2015quantum}
    Sun, Z., Yu, J., Wang, P., et al.: Quantum private comparison with a malicious third party. Quantum 431 Inf. Process. 14, 2125–2133 (2015)
    \bibitem{wang2016self}
    Wang, Q., Yu, C., Gao, F., et al.: Self-tallying quantum anonymous voting. Phys. Rev. A,  94, 022333 (2016) 
    \bibitem{ji2019quantum}
    Ji, Z.X., Zhang, H.G., Wang, H.Z., et al.: Quantum protocols for secure multi-party summation. Quantum Information Processing. 18(6), 168 (2019)
   \bibitem{zhang2014high}
   Zhang, C., Sun, Z.W., Huang, Y., et al.: High-Capacity Quantum Summation with Single Photons in Both Polarization and Spatial-Mode Degrees of Freedom. Int. J. Theor. Phys. 53(3), 933-941 (2014)
  \bibitem{liu2017novel}
  Liu, W., Wang, Y.B., Fan, W.Q.: An novel protocol for the quantum secure multi-party summation based on two-particle Bell states. Int. J. Theor. Phys. 56(9), 2783-2791 (2017)
   \bibitem{zhang2019quantum}
  Zhang, C., Razavi, M., Sun, Z., et al.: Quantum summation based on quantum teleportation. Entropy, 21(7), 719 (2019)
      \bibitem{chen10efficient}
   Chen, X.B., Xu, G., Yang, Y.X., Wen, Q.Y.: An efficient protocol for the secure multi-party quantum summation. Int. J. Theo. Phys. 49(11), 2793 (2010)
   \bibitem{zhang2015three}
   Zhang, C., Sun, Z.W., Huang, X., et al.: Three-party quantum summation without a trusted third party. Int. J. Quantum Inf. 13(02), 1550011 (2015)
   \bibitem{shi2016secure}
  Shi, R.H., Mu, Y., Zhong, H., et al.: Secure multiparty quantum computation for summation and multiplication.Sci. Rep. 6, 19655 (2016)\
  \bibitem{shi2017quantum}
  Shi, R.H., Zhang, S.: Quantum solution to a class of two-party private summation problems. Quantum Information Processing. 16(9), 225 (2017)
  \bibitem{yang2018secure}
    Yang, H.Y., Ye, T.Y.: Secure multi-party quantum summation based on quantum Fourier transform. Quantum Information Processing. 17(6), 129 (2018)
    \bibitem{li2005secure}
    Li, C.Y., Zhou, H.Y.,  Wang, Y., et al.: Secure Quantum Key Distribution Network with Bell States and Local Unitary Operations. Chin. Phys. Lett. 22(5), 1049-1052 (2005)
    \bibitem{li2006efficient}Li, C.Y., Li, X.H., Deng, F.G., Zhou, P., Liang, Y.J., et al.: Efficient Quantum Cryptography Network without Entanglement and Quantum Memory. Chin. Phys. Lett. 23(11), 2896 (2006)
   \bibitem{shor2000simple}
       Shor, P.W., Preskill, J.: Simple Proof of Security of the BB84 Quantum Key Distribution Protocol. Phys. Rev. Lett. 85(2), 441 (2000)
       \bibitem{gu2019improvement}
       Gu, J., Hwang, T.: Improvement on ``Secure multi-party quantum summation based on quantum Fourier transform''. arXiv:1907.02656 (2019).
\end{thebibliography}
\end{document}